\documentclass[aps,prl,twocolumn]{revtex4} %,preprint ,superscriptaddress
\usepackage{graphicx,bm,amssymb,amsmath}

\begin{document}
\title{Temperature dependence of the gain profile for THz quantum cascade lasers}
\author{Rikard Nelander}
\email[Email: ]{rikard.nelander@fysik.lu.se}
\author{Andreas Wacker}
\affiliation{Division of Mathematical Physics, Physics Department, Lund University, Box 118, 22100 Lund, Sweden}
\date{15 February 2008, to appear in Applied Physics Letters}

\begin{abstract}
We study the rapid decrease of peak gain in resonant-phonon THz Quantum Cascade Lasers with increasing temperature. The effect of various microscopic scattering processes on the gain profile as a function of temperature is discussed. We argue that increased broadening, primarily due to increased impurity scattering, and not diminishing population inversion, is the main reason for the reduction of peak gain.
\end{abstract}

\maketitle

The Quantum Cascade Laser (QCL) \cite{FaistScience1994} has proven to be a compact and robust coherent light source in the mid-IR region. The double-phonon resonance design has given the mid-IR QCL an operating range up to and above room temperature \cite{BeckScience2002}. QCLs in the THz-region are still restricted to low operating temperatures which limits practical applications. To this date, operation up to 164~K in pulsed and 117~K in cw mode have been demonstrated for THz QCLs \cite{WilliamsOE2005}. The common approach to stabilize population inversion by the double-phonon resonance has not been successful for THz laser \cite{WilliamsAPL2006}. Extending the operating region of these devices toward higher temperatures requires a quantitative understanding of the various microscopic temperature effects. Previously, theoretical studies have focused on the population inversion decrease with temperature \cite{IndjinAPL2003} and parasitic transport channels \cite{JirauschekPSSC2008}. Here we show that even for constant population inversion, the lasing performance can be strongly reduced with temperature in THz structures due to increased broadening of the gain profile.

\begin{figure}
\centering
\includegraphics[width=.99\columnwidth,keepaspectratio]{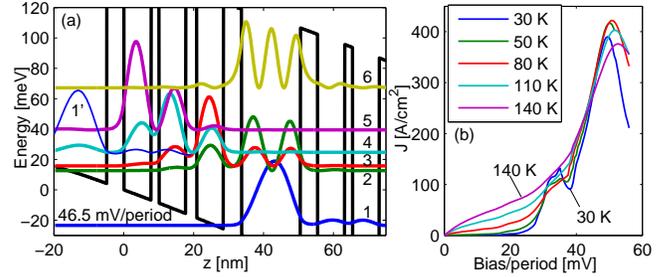}
\caption{(a) Conduction band profile with the modulus square of the six lowest Wannier-Stark states per period at the design bias of 46.5 mV/period. For clarity, the lowest state~(1') of the previous period is plotted as well. The lasing transition occurs between states~3 and~4. The barrier at $z \simeq 30$ nm is $\delta$-doped with a sheet density of $n = 2.25 \times 10^{10}$ cm$^{-2}$/period. (b) Calculated current-voltage relation. The experimental data shows the same shape and trend \cite{KumarPrivateCom}. However, the peak at 35~mV/period at low temperatures, corresponding to resonant tunneling from the injector state to the lower laser state (1'~-~3), is not observed and also the experimental current is 50~\% smaller, both at high and low bias.}
\label{states}
\end{figure}

In order to study these effects, the THz QCL from Ref.~\cite{KumarAPL2006} has been simulated. It is a four-well GaAs/Al$_{0.15}$Ga$_{0.85}$As design, see Fig.~\ref{states}(a), emitting at 1.9~THz (7.9~meV). Lasing was demonstrated up to 95~K in continuous wave operation and to 110~K in pulsed mode.

Our transport calculations are based on the Non-Equilibrium Green's Functions (NEGF) method~\cite{LeePRB2002, LeePRB2006}. This method can be seen as an extension of density matrix theory~\cite{SavicPRB2007, IottiPRB2005} to resolve broadening effects. We include electron-phonon, impurity, and interface roughness scattering \cite{TsujinoAPL2005,footnote1} within the self-consistent Born approximation, and electron-electron interaction within the mean-field approximation. Alloy scattering \cite{VasanelliAPL2006} is expected to be negligible for the GaAs/AlGaAs material system discussed here, since the well material is a binary compound. From the steady-state solution to the transport problem, see Fig.~\ref{states}(b), the gain is calculated according to the \emph{full theory} in Ref.~\cite{BanitAPL2005}. In our model the temperature enters in two ways: (\textit{i}) the occupation of phonon modes and (\textit{ii}) in the Debye approximation for the screening of ionized dopants. Throughout this letter, we refer to the lattice temperature $T$, which is typically larger than the heat-sink temperature \cite{VitielloAPL2005}. There is a monotonic dependence between both temperatures, which can be established by a detailed thermal modeling \cite{EvansJQE2006}, so that the general trends of performance are similar for both temperatures.

\begin{figure}
\centering
\includegraphics[width=.75\columnwidth,keepaspectratio]{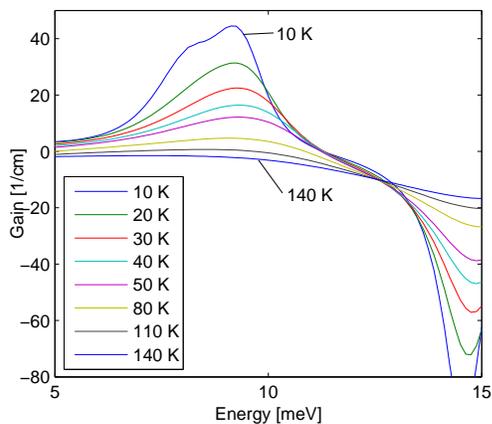}
\caption{Gain spectra for different temperatures and a bias drop of 46.5 mV/period.}
\label{gain}
\end{figure}

The obtained gain spectra are depicted in Fig.~\ref{gain}. The gain peak at 9~meV is strongly reduced with increasing temperature, in agreement with the observed vanishing of lasing at 110~K. A similar behavior is found for the absorption peak at 14~meV, corresponding to absorption between states 4~and~5.

\begin{figure}
\centering
\includegraphics[width=.99\columnwidth,keepaspectratio]{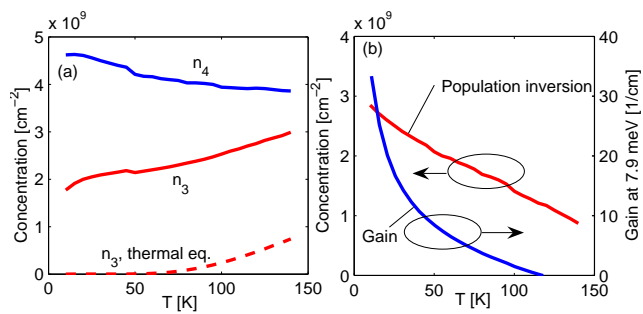}
\caption{(a) Subband population of the two lasing states as a function of temperature. The population of the lower laser state assuming thermal equilibrium with state~1 is also plotted (dashed) ($n_1 \simeq 1.3 \times 10^{10}$~cm$^{-2}$/period). (b) An approximate linear reduction of population inversion with temperature is found, while the gain at the lasing energy drops more rapidly. Estimating the  mirror losses, $\alpha_{\rm M} \sim$ 1.3 cm$^{-1}$, and the waveguide loss, $\alpha_{\rm W} \sim$ 4-5 cm$^{-1}$, according to Ref.~\cite{KumarAPL2006}, the maximum lasing temperature is below 100~K at this operation bias of 46.5~mV/period.}
\label{conc}
\end{figure}

Commonly, disappearing gain is related to vanishing population inversion. Fig.~\ref{conc}(a) shows that the population difference between the lasing states (3 and 4) indeed decreases with temperature. The occupation in state~3, assuming thermal equilibrium with respect to the heavily occupied state~1, is also depicted. The similar shape of the two curves suggests that thermal backfilling essentially stands for the reduction in population inversion. Since the lower laser state is separated from state~1 by approximately the optical phonon energy, this effect becomes relevant only at temperatures far above 100~K. Therefore, this decrease in population inversion is much less compared to the significant reduction of gain in the studied temperature range, see Fig.~\ref{conc}(b).

The excess reduction in peak gain can be related to a change in line shape. Indeed, Fig.~\ref{gain} shows that the gain transition is broadened with temperature, which reduces its peak value. The same holds for the absorption transition, which starts overlapping with the gain region at higher temperatures, causing an additional reduction of gain at the lasing energy.

The included temperature dependent scattering mechanisms are as follows:

(\textit{i}) \textit{Optical phonon} emission, the dominant scattering effect in most QCLs, is proportional to $n_B(E_{ \rm opt}) + 1$, where $n_B(E_{ \rm opt})$ is the Boltzmann-factor at the optical phonon energy. The phonon absorption rate is proportional to $n_B(E_{ \rm opt})$. At $T = 100$~K, $n_B(E_{ \rm opt}) \sim 10^{-2}$ for an optical phonon energy, $E_{ \rm opt}$, of $36.7$~meV. Thus, spontaneous optical phonon emission is by far the  dominant phonon process, which is temperature independent.

(\textit{ii}) \textit{Acoustic phonon} scattering contributes very little to the total scattering. Assuming elastic scattering and high temperature compared to the typical acoustic phonon, acoustic phonon scattering contributes less than 1\% to the total elastic scattering and can be neglected (see below).

(\textit{iii}) \textit{Impurity scattering} is strongly influenced by the screening of ionized dopants by electrons, which is a complex many-body problem where approximations are necessary \cite{LuAPL2006}. A common simple, but robust, approach is the Debye-approximation where the electrons contributing to screening are assumed to be in thermal equilibrium and obey Boltzmann statistics. This results in a temperature dependent screening length, \mbox{$\lambda^2_\mathrm{Debye} = \epsilon_s \epsilon_0 k_B T_e / e^2 n_\mathrm{3D}$}, where $n_\mathrm{3D}$ is the average electron concentration, and $\epsilon_s \epsilon_0$ is the dielectric constant of the sample. The idea is that a hot electron gas is less affected by the impurity potential and will, therefore, screen it less. The choice of the electron temperature $T_{e}$ and the three-dimensional electron density $n_{3D}$ is an intricate issue. For typical parameters of the sample considered here, the screening length is of the order of the period. Thus, electrons from all subbands are expected to contribute to the screening \cite{MeyerJAP1993}. As experiments for a similar structure show that the majority of electrons resides in subband 1, with an effective temperature close to the lattice temperature \cite{VitielloAPL2005}, we approximate $T_{e} = T$ and use the average electron density over the entire structure in $n_\mathrm{3D}$. At low temperatures compared to the Fermi energy $E_F$ screening is better described using the Thomas-Fermi approximation, \mbox{$\lambda^2_\mathrm{TF} = 2 \epsilon_s \epsilon_0 E_F / 3 e^2 n_\mathrm{3D}$.} In the calculations, the Thomas-Fermi model is used at temperatures below the crossing point of the screening lengths for the two models, {\it i.e.}, $k_B T = 2 E_F/3$ and vice versa.

We use a simple screening model in order to show the relevance of the mechanism. For a full quantitative description of screening in the nonequilibrium multisubband system of the QCL, a much more detailed study is desirable. Here, both aspects of the spatially inhomogeneous electron distribution \cite{LuAPL2006} as well as the frequency dependency of the dielectric function \cite{LeePRB1999}, in particular close to the optical phonon and plasma frequency, are of relevance. In this context it is also questionable if the distribution of the subbands are sufficiently well described by an effective temperature at all. All these aspect go, however, far beyond our present study.

\begin{figure}
\centering
\includegraphics[width=.99\columnwidth,keepaspectratio]{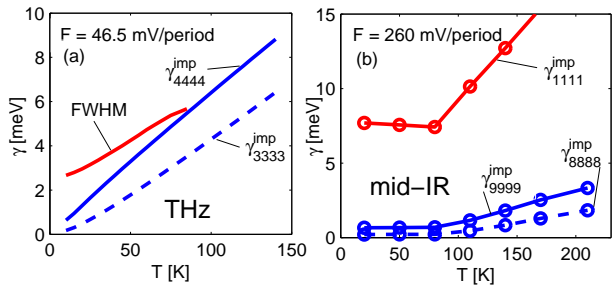}
\caption{$\gamma$-factors for intra-subband impurity scattering for the upper (solid) and lower (dashed) laser states for the THz (a) and mid-IR  device (b). The bright red curve in (a) shows the FWHM estimated for the gain peak in Fig.~\ref{gain}. In (b), the bright red curve displays $\gamma^{\rm imp}$ for the injector state (1). The kink in the $\gamma^\mathrm{imp}$ at 82 K is the cross-over from the Thomas-Fermi to the Debye screening model (10.7 K in the THz device), which is smoothed out in a more detailed calculation.}
\label{gamma}
\end{figure}

In order to quantify these scattering processes, the $\gamma$-factors, describing scattering matrix elements \cite{LeePRB2006,BanitAPL2005}, are plotted in Fig.~\ref{gamma}. The $\gamma$-factors relate to the scattering rate from state $\alpha$ to $\beta$ by $\Gamma_{\alpha  \rightarrow \beta} = \gamma_{\alpha \alpha \beta \beta}/\hbar$. The sum of the diagonal $\gamma$-factors, representing intraband scattering, for different scattering processes is a good measure of the energy width of a state. We observe a strong increase in the $\gamma^{\rm imp}$-factor with temperature for both lasing states.

Assuming high temperature compared to the typical acoustic phonon energy and linear phonon dispersion, the $\gamma$-factor in the elastic scattering approximation for acoustical phonon scattering reads
\begin{equation}
\gamma_{\alpha \alpha' \beta \beta'}^{\rm ac} = k_B T \frac{m^* D^2}{v^2 \rho \hbar^2} \int {\rm d} \! z \,  \psi^*_{\beta}(z)  \psi_{\alpha}(z)   \psi^*_{\beta'}(z) \psi_{\alpha'}(z),
\end{equation}
where $m^*$ is the effective mass, $D$ is the deformation potential, $v$ is the sound velocity, $\rho$ is the mass density, and $\psi_\alpha(z)$ is the wavefunction corresponding to Wannier-Stark state~$\alpha$. Using material parameters for GaAs from Ref.~\cite{VurgaftmanJAP2001}, we obtain $\gamma_{3333}^{\rm ac}$ = 0.024~meV at 100~K, suggesting that acoustic phonon scattering can indeed be neglected in this context.

In the THz device, interface roughness scattering is small compared to impurity scattering, $\gamma_{3333}^\mathrm{rough}$ = 0.12~meV and $\gamma_{4444}^\mathrm{rough} = 0.15$~meV, the same holds for the optical phonon emission $\gamma_{3311}^\mathrm{opt}$ = 0.45~meV from the lower laser state. Therefore, the FWHM shows a strong temperature dependence following the dominating impurity scattering, see Fig.~\ref{gamma}(a). Note that the FWHM is smaller than the sum of the individual widths due to correlations in the scattering environment \cite{BanitAPL2005}.

In mid-IR devices however, due to large conduction band offset between well and barrier material, interface roughness scattering can be stronger than impurity scattering. For the device presented in Ref.~\cite{SirtoriAPL1998}, for which our results are shown in Fig.~\ref{gamma}(b), we find $\gamma_{8866}^\mathrm{opt}$ = 1.36~meV, $\gamma_{8888}^{\mathrm{rough}} = 1.15$~meV, and $\gamma_{9999}^{\mathrm{rough}} = 11.7$~meV, which clearly dominates the total widths of the upper and lower laser states. Therefore, no strong temperature dependence of the gain profile is expected. An important aspect here is that the doping is located in the injector and is strongly screened by the large electron density, so that the lasing states are only weakly affected by impurity scattering. In contrast, states with strong spatial overlap with the dopants, e.g., the lowest injector state~1, exhibit large (and temperature dependent) broadening, see Fig.~\ref{gamma}(b).

In conclusion, our simulations show that lasing of THz QCLs can terminate at lattice temperatures where still a large portion of the population inversion remains. We attribute the excess gain reduction to increased scattering-induced broadening due to a reduction of screening with temperature. Thus, the common focus on population dynamics is likely to miss an additional mechanism of temperature dependent gain reduction.

The authors thank S.~Kumar and C.~Weber for helpful discussions and gratefully acknowledge financial support from the Swedish Research Council (VR).

\bibliographystyle{apsrev}

\begin{thebibliography}{23}
\bibitem[{foo()}]{footnote1}
\bibinfo{note}{In the simulation the average roughness height is set to $\eta$
  = 0.12 nm and correlation length $\lambda$ = 10 nm, similar to Ref. [14].}
\end{thebibliography}

\end{document}